# Predicting Melting point and Viscosity of Ionic Liquids Using New Quantum Chemistry Descriptors

A. Mehrkesh and A. T. Karunanithi

Center for Sustainable Infrastructure Systems, University of Colorado Denver, Denver, Colorado, 80217

**Abstract**

Ionic liquids (ILs) are an emerging group of chemical compounds which possess promising properties such as having negligible vapor pressure. These so called designer solvents have the potential to replace volatile organic compounds in industrial applications. A large number of ILs, through the combination of different cations and anions, can potentially be synthesized. In this context, it will be useful to intelligently design customized ILs through computer-aided methods. Practical limitations dictate that any successful attempt to design new ILs for industrial applications requires the ability to accurately predict their melting point and viscosity as experimental data will not be available for designed structures. In this paper, we present two new correlation equations towards the more precise prediction of melting point and viscosity of ILs solely based on the inputs from quantum chemistry calculations (no experimental data or simulation results are needed). To develop these correlations we utilized data related to size, shape, and electrostatic properties of cations and anions that constitutes ILs. In this work, new descriptors such as dielectric energy of cations and anions as well as the values predicted by an 'ad-hoc' model for the radii of cations and anions (instead of their van der waals radii) were used. An enormous form of correlation equations constituent of all different combinations of descriptors (as the inputs to the model) were tested. The average relative errors were measured to be 3.16% and 6.45% for the melting point, $T_m$, and ln(*vis*), respectively.

## 1. Introduction

Ionic liquids (ILs) are an emerging group of chemicals having interesting properties from negligible volatility, and high thermal stability to wide electrochemical window, and high solvency power. The aforementioned characteristics make ionic liquids a promising candidate for multitude of applications. Ionic liquids are mainly composed of large organic cations and organic or inorganic anions. Physical properties of ionic liquids can significantly vary from their sister compounds, ionic salts. The properties of ionic salts can be completely and solely attributed to their ionic nature since strong ionic bonds hold their particles together. Ionic salts are mostly made of small monoatomic ions in the close vicinity of one and other in their lattice crystal network. The lattice energy of crystalline compounds is proportional to the inverse of the distance between their lattice points. The ionic bonds in salts are very strong due to the relatively short distance between the small ions contributing to high values of melting point. On the contrary, ionic liquids are made up of larger multi-atomic cations and anions resulting in weaker ionic bonds. This explains the considerably low melting point and viscosity of ionic liquids compared to those of ionic salts. The multi-atomic nature of cations and anions in ionic liquids is a great opportunity for researchers to fine tune their properties with customizing them for different applications. In ILs, cations, and occasionally anions, are composed of alkyl side chain groups ($CH_2$, $CH_3$, etc.) which can be accompanied with a number of functional groups (OH, $NH_2$, COOH, etc.) to functionalize ionic liquids for different applications. A vast number of ILs (estimated to be about $10^{14}$)[1] can be potentially synthesized through distinct combinations of different cation-cores, alkyl groups, functional groups, and anions. Careful evaluation of experimental data from literature on the physical and thermodynamic properties of ILs shows that substituting functional groups can drastically alter a property of interest in ionic liquids.

Despite the fact that the ionic bonds in ILs are relatively weak, their properties can still be attributed to their ionic nature as even a weak ionic bond is still stronger than any other type of intermolecular forces. Crystal network of ionic materials (e.g. ionic liquids) are made of cations and anions held together by electrostatic attraction. The ionic force between charged particles is directly proportional to the charge of each particle and inversely to the distance between them. The larger the cations and anions are, the weaker the ionic bonds between them will be. This is due to the fact that by increasing the distance between two ions the electrostatic attraction, which holds them together, is reduced.

Computer-aided optimization methods can help in designing optimal ionic liquids suitable for different applications from solvent extraction to thermal energy storage.[2,3] ILs are generally salts that are in liquid phase below an arbitrary temperature, commonly 100°C. Therefore, several ILs are in solid state at room temperature. When it comes to the melting point of ILs, those with significantly lower melting points or "room temperature ionic liquids [RTILs]" ($T_m$<25°C), are of great interest to researchers seeking new industrial applications. One main reason for the desirability of low melting point of ILs is the fact that they are being considered as candidates (solvents or absorbents) for separation tasks involving selective dissolution of solutes in gas (e.g. $CO_2$), liquid (e.g. toluene), or solid (e.g. cellulose) states. They are also widely considered as liquid solvents to promote chemical reactions.[4,5] From a practical point of view, for an IL to be used as an industrial solvent it needs to be transported (pumped) across multiple unit operations and therefore it must be in the liquid phase.

Another significant barrier towards commercialization of IL-based applications is their high viscous characteristics due to their ionic nature (existence of relatively strong ionic bonds) resulting in transportation difficulties. This necessitates the use of powerful pumping equipment and efficient process equipment to handle viscous fluids. Therefore, identification and customization of ILs that have relatively low viscosity and appropriate melting point will greatly aid in their accelerated discovery and commercialization. Several models exist for prediction of melting point and viscosity of ILs. In general, these models are either Quantitative Structure Property Relationship (QSPR) correlations[6-8], group contribution (GC) type[9-16], or molecular dynamics (MD) simulations.[17-23] The main drawback of the QSPR and GC models is the narrow restrictions in choice of applicable compounds while the main limitation of MD simulations is the high computational demands and the need for starting geometry.[24] In this paper, we present two empirical correlations to predict melting point and viscosity of ILs in a way that does not need experimental input or tedious simulations, but rely on inputs from simple quantum chemical calculations.

## 2. Methods

Widely available data and research findings related to melting point and viscosity of salts explicitly shows that there is a meaningful relationship between the magnitude of the two properties and the lattice energy of ionic bonds. Generally, in an ionic compound, size (volume and area), shape (sphericity), molecular weight, and dielectric energy of ions, play important roles in determining the strength of an ionic bond. Normally, larger size of cations and anions, results in longer distances between these ions in the crystal lattice, making the ionic bond weaker. On the other hand, the shape of the ions is also important as they are better packed together when they are symmetrical in nature. It has been suggested that asymmetry of ions in an ionic compound, most likely, will decrease the melting point since the ions are more loosely connected and can be easily separated from each other (by applying lower amount of energy). When we studied the size of cations and anions of a wide variety of ionic liquids and compared with their melting points and viscosities, we found that there were several ILs that violated these general trends, i.e. ILs with relatively large cations and anions did not necessarily have low melting point or viscosity. One reason could be that in ionic liquids, ionic bonds are not the only intermolecular forces responsible for holding the particles together and other types of forces such as hydrogen bonding and polar-polar forces also come into play. Therefore we focused on using size, shape, and electrostatic properties to develop new correlations for the prediction of ILs' melting points and viscosities. In order to better account for deviations discussed above we included, in addition to the three descriptors, several other quantum chemical descriptors to refine the correlations to cover a wide variety of ionic liquids. Another challenge was that for several ILs, there were multiple experimental values reported in literature for these two physical properties. This inconsistency was especially observed in the case of melting points which could be due to the existence of impurities or different crystal network systems. In this study, we did not include ILs that had inconsistent experimental values, during the development of the correlations. Ionic liquids selected for this study were all 1:1 (one cation and one anion) with delocalized charges. These types of ILs are normally able to avoid crystallization and form glasses compounds far below room temperature.[25]

Currently, the most widely used approach to predict the melting point of ILs is quantitative structure-property relationship (QSPR) models, mostly combined with artificial neural networks (ANNs).[25] In this approach, there is a reasonably good correlation between actual and predicted melting points within a standard deviation of less than 10°C.[25] The limited availability of experimental data on physical properties of ILs is the main drawback of constructing good QSPR models. In recent years, simulations with molecular dynamics (MD) have evolved to study the behavior of ILs. The quality of these simulations strongly depends on the employed force fields. Several groups have tuned them specifically for ILs, while others have modified previously existing force fields. For example, Alavi and Thompson[25] have used MD simulations to predict the melting temperature of [$C_2$MIm][$PF_6$]. The demanding simulation indicated a melting point that was approximately 43°C too high.[25]

Quantum chemistry data related to the constituents of the ILs, i.e. cations and anions, were obtained using TURBOMOLE software which is a powerful, general-purpose Quantum Chemistry (QC) program, for ab-initio electronic structure calculations.[26] This software allows for accurate prediction of cluster structures, conformational energies, excited states, and dipoles that can be utilized for different purposes. When a chemical compound, in our case a cation or an anion, is simulated using TURBOMOLE, it can be exported as a Cosmo file, which can be used in COSMOtherm software.[26] COSMOtherm is a universal tool, which combines quantum chemistry (QC) and thermodynamic laws to calculate different properties of chemicals and solvents.[27] This method can estimate the chemical potential of different molecules (in pure or mixed forms) at different temperatures. In contrast to other methods, COSMOtherm can predict thermodynamic properties of compounds as a function of concentration and temperature by equations, which are thermodynamically consistent.[27] In the this study, for all the cations and anions, their corresponding COSMO files were imported from COSMO*baseIL* database (Version 1501), provided by the Cosmologic Company, in which the compounds geometries are optimized, the ground state energies are calculated, and the corresponding COSMO and gasphase files are generated all using TZVP (BP_TZVP_C30_1501) basis set. In the next step, the computational chemistry information of the selected cations and anions, required for the development of the correlative equations, were collected from their COSMO files. These data were either used directly (without any changes) or as a combination with other data (as the secondary descriptors) as the inputs to the correlative equations utilized to predict the melting point and viscosity of ILs.

## 3. Results and Discussion

Molar mass of cation ($MW_C$), molar mass of anion ($MW_A$), volume of anion $[A^o]^3$ ($Vol_A$), volume of cation $[A^o]^3$ ($Vol_C$), Area of anion $[A^o]^2$ ($Area_A$), Area of cation $[A^o]^2$ ($Area_C$), dielectric energy of cation ($Di_C$), dielectric energy of anion ($Di_A$), symmetrical values of ions ($\sigma$), density $[\frac{kg}{m^3}]$ ($\rho$) and the average distance between a cation and an anion in the network $[A^o]^3$ ($R_t$) were selected as descriptors to develop the proposed empirical correlations. Accurate measurements of volume, surface area and ionic radii of cations and anions are only possible through X-ray diffraction method. When X-ray data is unavailable correlative or approximation methods (e.g. van der Waals model) need to be used to estimate the characteristics that are related to the size and shape of ions. X-ray diffraction data is very sparse for the case of ILs and since we are interested in development of a universal correlation covering a wide range of IL structures, we have to rely on predictive data. Typical cations and anions of ILs, do not usually have spherical shapes, so it is necessary to estimate their ionic radii using correlative approaches. We selected ions for which the value of the experimental ionic radii were available. Next, for those selected ions, we also estimated the ionic radii using van der Waals model where all cations and anions are assumed to have a spherical shape. The volumes of corresponding ions were estimated using COSMOtherm software and were converted to ionic radii using Eqn.1.

$$R_{Calc} = \frac{3}{4} \times \frac{1}{\pi} \times Vol^{1/3} \qquad \text{Eqn.1}$$

A linear correlation between the experimental values of the cations and anions radii and the corresponding values of their van der Waals radii is shown in Figure 1. As it can be seen from Figure 1, the actual and predicted ionic radii are perfectly correlated through a linear relationship ($R^2$=0.98622). The corresponding linear fit is shown in Eqn.2.

$$R_{Pred} = 0.48601 \times R_{Calc} - 0.01725 \qquad \text{Eqn.2}$$

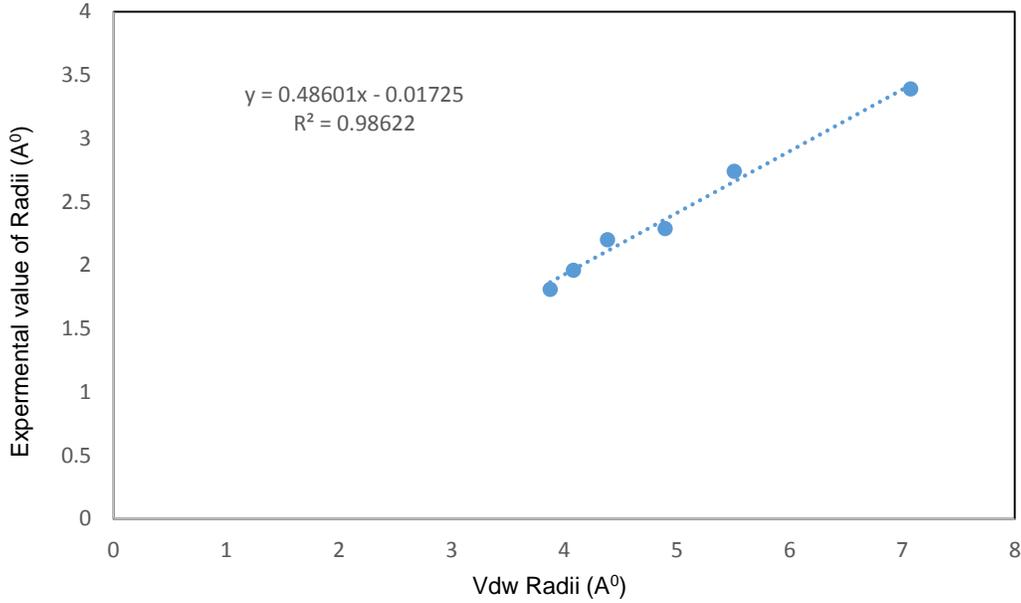

**Figure 1.** A correlation between the van der Waals (VdW) and actual radii of ions in ILs

Although, there were only few experimental data points available for use in this correlation, we see that this data covers a wide range of VdW ionic radii from 3.8 to 7.5 A$^0$. Therefore, we were comfortable in using this linear relationship to estimate the unknown values of ionic radii of other cations and anions considered in this study. First, the ionic radius of each cation or anion was calculated using Eqn.1 from the volume calculated by COSMOtherm software. This value was corrected using Eqn.2 to arrive at the predicted ionic radii. The distance between anions and cations, $R_t$, were estimated as the sum of ionic radii of cations and anions present in the crystal network of ILs. The symmetrical value of ionic liquids, σ, were calculated using the sphericity of cations and anions. Sphericity is a measure of how spherical an object is and can be estimated using Eqn.3.

$$Sph = \frac{\pi^{\frac{1}{3}}(6V_p)^{\frac{1}{3}}}{A_p} \qquad \text{Eqn.3}$$

where $V_p$ and $A_p$ are the volume and surface area, respectively.

Further, the symmetrical value of ionic liquids were calculated using Eqn.4.

$$\sigma = \sqrt{Sph_C * Sph_A} \qquad \text{Eqn.4.}$$

where $Sph_C$ and $Sph_A$ are the sphericity of cations and anions in the IL of interest, respectively.

## 3.1 Melting point

Experimental melting point data of several ILs covering a wide range of categories (different type of cation cores, anions, functional groups etc.), were gathered from the literature.[28-46] A multivariate correlation with inputs based on quantum chemistry descriptors, and the output of the experimental values of melting point, $T_m$, was performed utilizing Eureqa software, a data analysis tool developed by Nutonian, Inc. Out of the 37 experimental melting point data, 17 (i.e. 45% of the data points) were chosen as part of the validation set (depicted in green in Figure 2), which were not used in the process of the model development.

The experimental data along with the correlated multivariate trend line, that was developed to predict the melting point of ILs as a function of their QC parameters, is shown in Figure 2. It can be seen from Figure 2 that the trend line strongly captures the fluctuations in the melting point data.

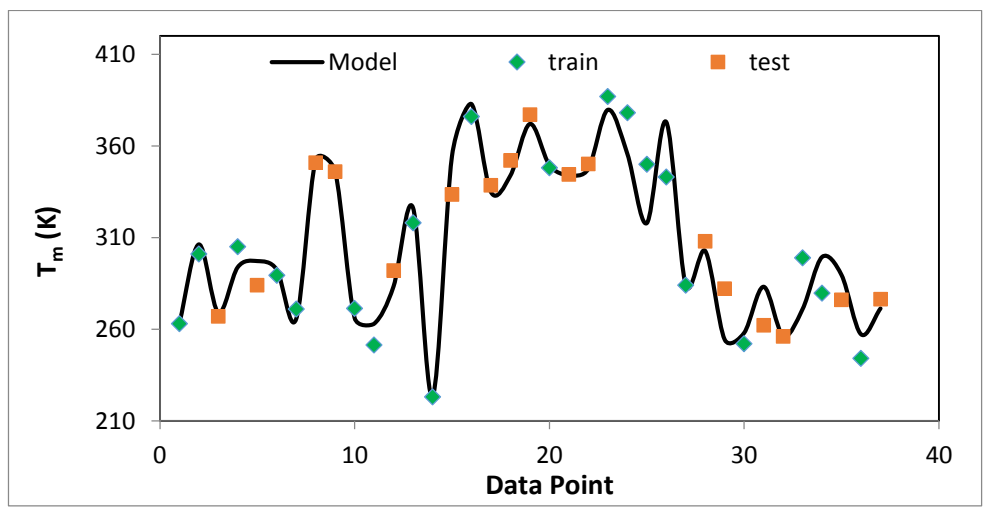

**Figure 2.** Experimental vs. model predicted melting point data (black-Training set; Green-Validation set)

Figure 3 presents a comparison between the experimental (observed) melting point data and the corresponding predictions based on the developed correlation.

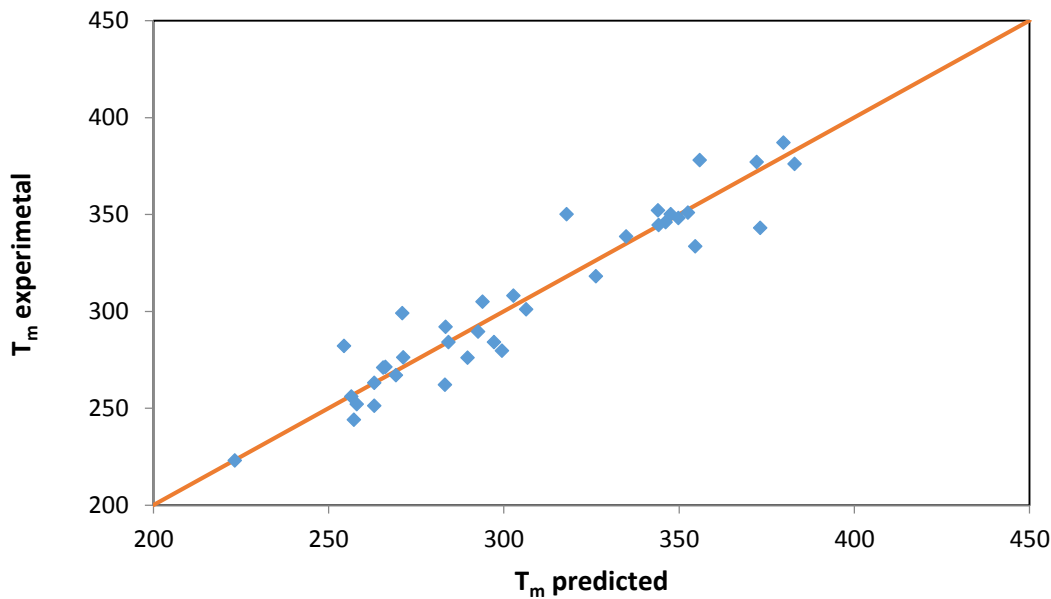

**Figure 3.** Goodness of the model predictions for the melting point of ILs

The best mutivariative correlation (having highest $R^2$) for predicting the melting point of ILs, shown in Eqn.5, uses the QC descriptors of the cation and anion radii, density of ionic liquids, symmetrical value, molecular weight of cation, and the dielectric energy of anion.

$$T_m = a\,(R_A) + b\,(\rho) + c\,(R_C)(\sigma) - d - e(MW_C) - f(Di_A) \tag{5}$$

a=11.38, b=0.05413, c=196.6, d=434.3, e=0.649, f=1661

Table 1 lists the characteristics related to the selected predictive empirical correlation.

**Table 1.** Model Accuracy for prediction of melting point

| Parameter | Value-All data | Value-Test data | Value-Train data |
|---|---|---|---|
| $R^2$ (Goodness of Fit) | 0.91 | --- | --- |
| Maximum Relative Error (%) | 9.87 | 9.87 | 9.41 |
| Avg. Relative Error (%) | 3.16 | 2.77 | 3.48 |

**3.2 Viscosity**

Experimental data on the viscosity of several ILs at different temperatures, were gathered from the literature.[46-65] Once more, a multivariate correlation with several quantum chemistry descriptors along with temperature as input parameters, and ln (*viscosity*), as the output parameter, was developed using Eureqa software.

In this case, we used 131 viscosity data points, out of which, 83 were chosen as part of the validation set (depicted in green in Figure 4), which again were not used for model development purposes. The experimental viscosity data points and the correlated multivariate trend line that was developed to predict the viscosity of ILs as a function of QC descriptors, is shown in Figure 4. It can be seen from Figure 4 that, the trend line strongly captures the fluctuations in the viscosity of selected ILs.

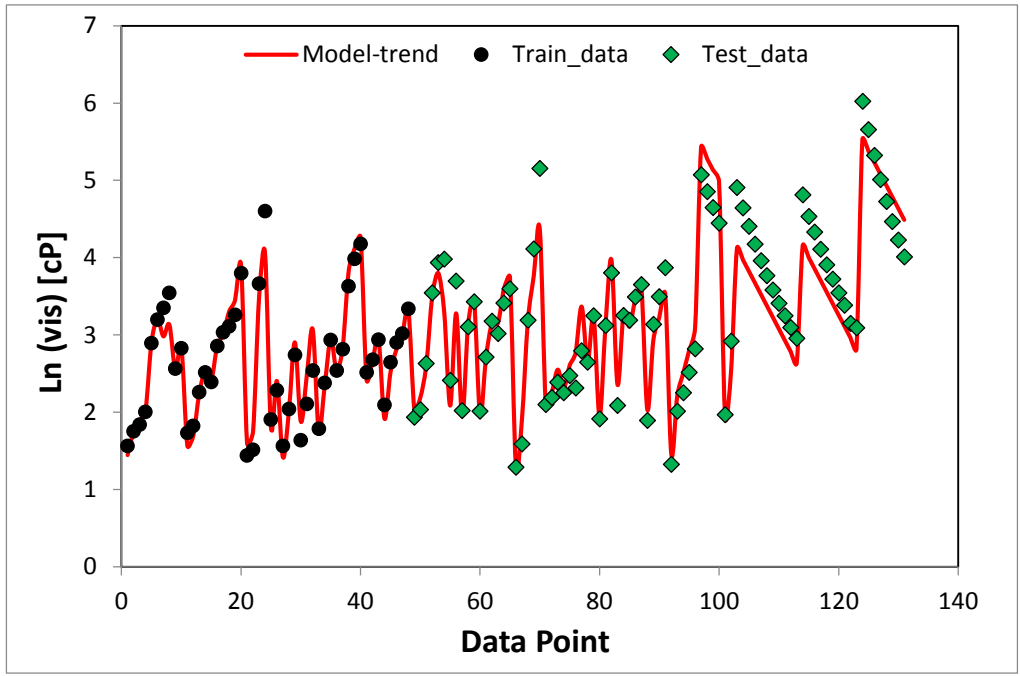

**Figure 4.** Experimental vs. model predicted viscosity data (black-Training set; Green-Validation set)

Figure 5 presents a comparison between the experimental viscosity data and the corresponding predictions based on the developed correlation.

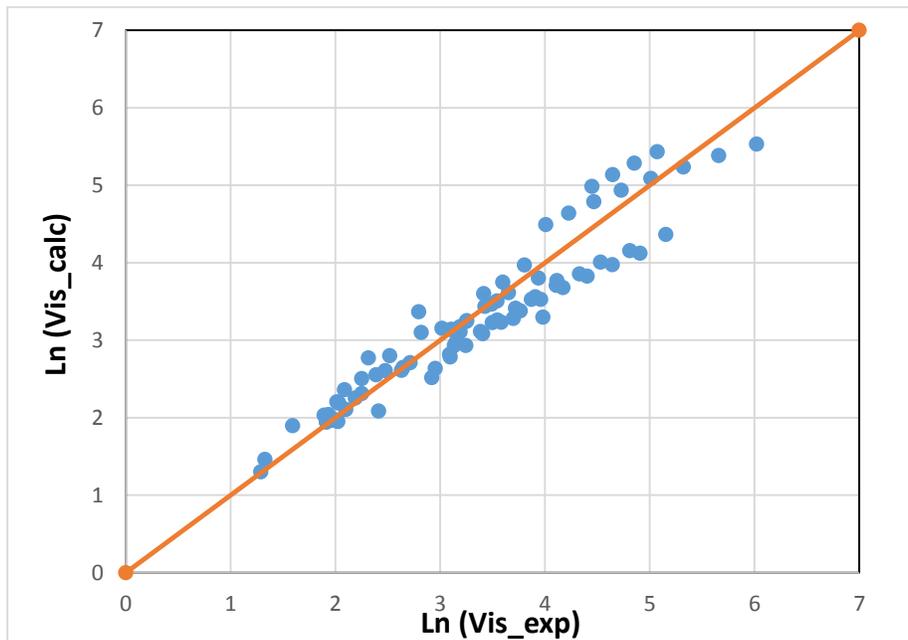

**Figure 5.** Goodness of the model for prediction of viscosity of ILs-test data only

The best multivariate correlation (having highest $R^2$) for viscosity, shown in Eqn.6, uses the descriptors of the symmetrical value, distance between anions and cations, anion volume and surface area, temperature and the dielectric energy of anion and cation.

$$\ln(vis) = a\,(\sigma) + b\,(R_t) + c\,(Vol_A) - d - e\,(Area_A) - f\,(T) - g\,(Di_A) - h\,(Di_C) \qquad \text{Eqn.6}$$

a=16.513, b=2.2179, c=0.00892, d=15.0073, e=0.02686, f=0.02975, g=15.8297, h=48.1367

Table 2 lists the characteristics related to the empirical model used to predict the viscosity of ILs as a function of QC parameters and temperature.

It is noteworthy that the average error in the prediction of viscosity of different classes of ionic liquids (i.e. ionic liquids with different cation head groups at a wide range of temperature) using the semi-empirical model of this study was calculated to be 6.45%. This shows a significant improvement in the accuracy of the prediction compared to the model used in Preiss et al.[25] (which has ~18% absolute error for the same group of ionic liquids) study which is currently incorporated in COSMOTherm software.

**Table 2.** Model Accuracy for prediction of viscosity

| Parameter | Value-All data | Value-Test data | Value-Train data |
|---|---|---|---|
| $R^2$ (Goodness of Fit) | 0.91 | --- | --- |
| Maximum Relative Error (%) | 20.78 | 20.51 | 20.78 |
| Avg. Relative Error (%) | 6.45 | 7.40 | 4.81 |

## 4. Comparison

There have been several studies focusing[9-16] on prediction of the physical properties of ionic liquids using group contribution (GC) approaches. In these methods, a value is attributed to each structural group (a cation head group, a side chain group attached to the cation and an anion). Optimization frameworks are used to regress experimental data points on the physical properties of chemicals, in this case ionic liquids, to equations comprising the weighted summation of the contribution of structural groups (i.e. number of each structural group multiply by its corresponding contribution). GC methods are limited to the structural groups for which experimental data are available. Therefore, these methods, even if accurate, can only be used for the structural groups which were available in the initial set of data used to develop the model. This study is not to invalidate the group contribution models used predict viscosity and melting point of ionic liquids since these methods can be very useful in trend analysis in which the effect of different structural groups on physical properties can be studied. Only it is noteworthy that models errors in this study for prediction of two of the most important physical properties of ionic liquids (melting point and viscosity) are comparable to those achieved by group contribution models[9-16] (i.e. less than 10% absolute error). The main difference is that the models in this study are solely *ab initio* meaning that there is no need for experimental data and they can be used to predict properties of all classes of ionic liquids even those for which no experimental data is available thereby GC models cannot be applied.

## 5. Conclusion

In this study, we present two multivariate empirical correlations to predict the melting point and viscosity of ILs with relatively high accuracy. Prediction of these two properties of ILs is critical as they have a significant influence on the practical applicability of ILs for industrial applications. The input parameters in the case of melting point (Tm) prediction were all quantum chemistry (QC) descriptors and in the case of viscosity were QC descriptors and temperature. These correlations can be utilized as long as the structure of a new ionic liquid is known and the quantum chemistry calculations can be performed to simulate the IL and estimate the descriptors related to size, shape, electrostatic energies, and the polarity which can be used to predict the macroscopic properties, melting point and viscosity. The correlative model developed for the prediction of viscosity of ionic liquids can be used to predict transport properties, such as diffusion coefficient, of novel ionic liquids which can be further used to conduct environmental fate and transport modeling of these ionic liquids using the framework described in our previous work.[66]

**Acknowledgement**
This material is based upon work supported by the United States National Science Foundation (CAREER Program) under Grant No. 1151182.